\documentstyle[preprint,prb,aps]{revtex}

\begin{document}

\begin{titlepage}

\title
{Phase interference of spin tunneling in an arbitrarily directed magnetic field}

\author{Rong L\"{u}\footnote 
{Electronic address: rlu@castu.tsinghua.edu.cn}, Jia-Lin Zhu, 
Yi Zhou, and Bing-Lin Gu} 
\address{Center for Advanced Study, 
Tsinghua University, Beijing 100084, People's Republic of China
}
\date{\today}
\maketitle
\begin{abstract}
We present an exact analytic study on 
the topological phase interference effect
in resonant quantum tunneling
of the magnetization between degenerate excited levels for biaxial ferromagnets
in an arbitrarily directed magnetic field.
We show that the topological phase interference effect depends
on the orientation of the field distinctly.
The transition from classical to quantum behavior is also discussed.

\noindent
{\bf PACS number(s)}:  75.45.+j, 75.10.Jm, 03.65.Bz
\end{abstract}

\end{titlepage}

During the last decade, there has been great interest in the problem of
quantum tunneling in nanometer-scale magnets.\cite{1} One notable subject is
that the topological Wess-Zumino (or Berry) phase\cite{2} can lead to
remarkable spin-parity effects.\cite{3,4,5,6,7,8,9,10,11,12,13,14,15} It was
theoretically shown that the tunnel splitting is suppressed to zero for
half-integer total spins in single-domain biaxial ferromagnetic (FM)
particles due to the destructive interference of the Berry phase between two
tunneling paths of opposite windings.\cite{3} However, the interference is
constructive for integer spins, and hence the splitting is nonzero. While
spin-parity effects are sometimes be related to the Kramers degeneracy, they
typically go beyond the Kramers theorem in a rather unexpected way.\cite{4,5}
The auxiliary particle method was proposed to study the spin-parity effects
in one model of a single large spin subject to the external and anisotropy
fields.\cite{6} Recently, the spin-phase interference and quantum
oscillation effects were studied extensively in FM particles in the presence
of a magnetic field, with the field along either the hard,\cite{4,7} easy,%
\cite{8} or medium axis.\cite{9} Similar spin-parity effects were also found
in antiferromagnetic (AFM) particles,\cite{10,11} and in the quantum
propagation of Bloch walls in quasi-one-dimensional ferromagnets\cite{12}
and antiferromagnets.\cite{13} By applying the effective Hamiltonian
approach, the effect of magnetic field and quantum interference of the
magnetization vector (or the N\'{e}el vector) were studied in FM (or AFM)
particles with different symmetry.\cite{14}

Experiment on the molecules Fe$_8$ showed a direct evidence of the role of
the topological phase in spin dynamics.\cite{15} Recent theoretical and
experimental studies include the spin tunneling in a swept magnetic field,%
\cite{16} the thermally activated resonant tunneling based on the
perturbation theory\cite{17} and the exact diagonalization,\cite{18} the
discrete WKB\ method and a nonperturbation calculation,\cite{19} the
non-adiabatic Landau-Zener model,\cite{20} the calculation based on exact
spin-coordinate correspondence,\cite{21} and the effects caused by the
higher order term and the nuclear spins on the tunnel splitting of Fe$_8$.%
\cite{22}

It is noted that the previous studies on FM spin-parity effects\cite
{4,7,8,9,14} were mostly focused on the phase interference between two
opposite winding ground-state tunneling paths. Moreover, the previous works%
\cite{4,7,8,9} have been confined to the condition that the magnetic field
be applied along the easy, medium, or hard axis, separately. In this paper
we study the topological phase interference effects for single-domain
biaxial FM particles in an arbitrarily directed magnetic field. {\it Our
study provides a nontrivial generalization of the Kramers degeneracy for
equivalent double-well system to coherently spin tunneling at ground states
as well as low-lying excited states for FM\ system with asymmetric twin
barriers caused by the arbitrarily directed magnetic field.} Integrating out
the momentum in the path integral, the spin tunneling problem is mapped onto
a particle moving problem in one-dimensional periodic potential $U\left(
\phi \right) $. By applying the periodic instanton method, we obtain exactly
the splittings between degenerate excited levels of neighboring wells. $%
U\left( \phi \right) $ is regarded as a one-dimensional superlattice with a
periodically recurring asymmetric twin barriers. The general translation
symmetry results in the energy band structure, and the low-lying energy
level spectrum is obtained by using the Bloch theorem. Our result shows that
the excited-level splittings depend significantly on the parity of total
spins of FM particles, and {\it this spin-parity effect depends on the
orientation of the field distinctly}. The transition from quantum to
classical behavior is also studied, and the second-order phase transition is
shown.

For a spin tunneling problem, the tunneling rate is determined by the
Euclidean transition amplitude 
\begin{equation}
K_E=\left\langle f\right| e^{-{\cal H}T}\left| i\right\rangle =\int D\Omega
\exp \left( -S_E\right) ,  \eqnum{1}
\end{equation}
where $D\Omega =\sin \theta d\theta d\phi $, and the Euclidean action is 
\begin{equation}
S_E\left( \theta ,\phi \right) =\frac V\hbar \int d\tau \left[ i\frac{M_0}%
\gamma \left( \frac{d\phi }{d\tau }\right) -i\frac{M_0}\gamma \left( \frac{%
d\phi }{d\tau }\right) \cos \theta +E\left( \theta ,\phi \right) \right] . 
\eqnum{2}
\end{equation}
$M_0V=\left| \overrightarrow{M}\right| V=\hbar \gamma S$, where $V$ is the
volume of the particle, $\gamma $ is the gyromagnetic ratio, and $S$ is the
total spins. The first two terms in Eq. (2) define the topological
Wess-Zumino term which arises from the nonorthogonality of spin coherent
states.\cite{3} The total derivative has no effect on the classical
equations of motion, but is crucial for the spin-parity effects.

The system of interest has the biaxial symmetry, with $\widehat{x}$ being
the easy axis, $\widehat{y}$ being the medium axis, and $\widehat{z}$ being
the hard axis. The magnetic field is applied in the $ZY$ plane, at an angle
in the range of $\frac \pi 2\leq \theta _H\leq \pi $. Now $E\left( \theta
,\phi \right) $ in Eq. (2) is 
\begin{equation}
E\left( \theta ,\phi \right) =K_{\bot }\cos ^2\theta +K_{\Vert }\sin
^2\theta \sin ^2\phi -M_0H_y\sin \theta \sin \phi -M_0H_z\cos \theta , 
\eqnum{3}
\end{equation}
where $H_y=H\sin \theta _H$, $H_z=H\cos \theta _H$, $K_{\Vert }$ and $%
K_{\bot }$ are the longitudinal and the transverse anisotropy coefficients
satisfying $K_{\bot }\gg K_{\Vert }>0$. As $K_{\bot }\gg K_{\Vert }>0$, the
deviations of $\theta $ about $\theta _0$ are small. Introducing $\theta
=\theta _0+\alpha $, $\left| \alpha \right| \ll 1$, Eq. (3) reduces to 
\begin{equation}
E\left( \alpha ,\phi \right) =K_{\bot }\sin ^2\theta _0\alpha ^2+K_{\Vert
}\sin ^2\theta _0\left( \sin \phi -\sin \phi _0\right) ^2,  \eqnum{4}
\end{equation}
where $\cos \theta _0=\frac{M_0H_z}{2K_{\bot }}$, $\sin \phi _0=\frac{M_0H_y%
}{2K_{\Vert }\sin \theta _0}=\frac{h\sin \theta _H}{\sqrt{1-\left( \lambda
h\cos \theta _H\right) ^2}}$, $\lambda =\frac{K_{\Vert }}{K_{\bot }}$, $h=%
\frac H{H_0}$, and $H_0=\frac{2K_{\Vert }}{M_0}$. Performing the Gaussian
integration over $\alpha $, we obtain the transition amplitude as 
\begin{equation}
K_E=\exp \left[ -iS\left( 1-\cos \theta _0\right) \left( \phi _f-\phi
_i\right) \right] \int d\phi \exp \left\{ -\int d\tau \left[ \frac 12m\left( 
\frac{d\phi }{d\tau }\right) ^2+U\left( \phi \right) \right] \right\} , 
\eqnum{5}
\end{equation}
with $m=\frac{\hbar S^2}{2K_{\bot }V}$, and $\hbar U\left( \phi \right)
=K_{\Vert }V\sin ^2\theta _0\left( \sin \phi -\sin \phi _0\right) ^2$. Since
the configuration space of this problem is a circle, calculations are
restricted to the first twin barrier at $\phi =\frac \pi 2$ and $\frac{3\pi }%
2$. We use $A$ to denote the instanton passing through the small barrier at $%
\phi =\frac \pi 2$ with the height $\hbar U_S=K_{\Vert }V\sin ^2\theta
_0\left( 1-\sin \phi _0\right) ^2$, and $B$ through the large barrier at $%
\phi =\frac{3\pi }2$ with the height $\hbar U_L=K_{\Vert }V\sin ^2\theta
_0\left( 1+\sin \phi _0\right) ^2$. Correspondingly, there are two kinds of
anti-instantons: $A^{-}$ and $B^{-}$.

The periodic instanton configuration $\phi _p$ which minimizes the Euclidean
action in Eq. (5) at an energy $E>0$ satisfies the equation of motion 
\begin{equation}
\frac 12m\left( \frac{d\phi _p}{d\tau }\right) ^2-U\left( \phi _p\right) =-E.
\eqnum{6}
\end{equation}
Then we obtain the periodic instanton $A$ solution as 
\begin{equation}
\sin \phi _A=\frac{1-\xi _1\text{sn}^2\left( \omega _1\tau ,k_1\right) }{%
1+\xi _1\text{sn}^2\left( \omega _1\tau ,k_1\right) },  \eqnum{7}
\end{equation}
corresponding to the transition of $\phi $ from $\arcsin \alpha $ to $\pi
-\arcsin \alpha $, where $\alpha =\sin \phi _0+\sqrt{\frac{\hbar E}{K_{\Vert
}V\sin ^2\theta _0}}$. sn$\left( \omega _1\tau ,k_1\right) $ is the Jacobian
elliptic sine function of modulus $k_1$, where $k_1^2=\frac{\left( 1-\alpha
\right) \left( 1+\beta \right) }{\left( 1+\alpha \right) \left( 1-\beta
\right) }$, $\beta =\sin \phi _0-\sqrt{\frac{\hbar E}{K_{\Vert }V\sin
^2\theta _0}}$, $\xi _1=\frac{1-\alpha }{1+\alpha }$, $\omega _1=\frac{%
\omega _0}{g_1}$, $\omega _0=2\frac{\sqrt{K_{\bot }K_{\Vert }}V}{\hbar S}%
\sin \theta _0$, and $g_1=\frac 2{\sqrt{\left( 1+\alpha \right) \left(
1-\beta \right) }}$. The classical action or the WKB\ exponent in the tunnel
splitting is obtained by integrating the Euclidean action in Eq. (5) with
the above periodic instanton solution. The result for instanton $A$ is $%
S_A=W_A+2E\beta $, where 
\begin{equation}
W_A=4m\omega _1\left[ E\left( k_1\right) +\frac{\left( k_1^2-\xi _1\right) }{%
\xi _1}K\left( k_1\right) +\frac{\left( \xi _1^2-k_1^2\right) }{\xi _1}\Pi
\left( k_1,\xi _1\right) \right] .  \eqnum{8}
\end{equation}
$K\left( k_1\right) $, $E\left( k_1\right) $, and $\Pi \left( k_1,\xi
_1\right) $ are the complete elliptic integral of the first, second, and
third kind.

Now we discuss briefly the calculation of the tunnel splitting. For a
particle with mass $m$ moving in a smooth double-well-like one-dimensional
potential $U\left( x\right) $, the instanton approach gives the ground-state
tunnel splitting as\cite{23} 
\begin{equation}
\Delta E_0=2C\left( \frac{W_0}{2\pi }\right) ^{1/2}\exp \left( -W_0\right) ,
\eqnum{9}
\end{equation}
where $W_0$ is the classical action for the ground-state tunneling, 
\begin{equation}
C=\left\{ \frac{\det \left( -\partial _\tau ^2+\omega ^2\right) }{%
\det^{\prime }\left[ -\partial _t^2+U^{\prime \prime }\left( x_{cl}\left(
\tau \right) \right) /m\right] }\right\} ^{1/2}  \eqnum{10}
\end{equation}
is the ratio of fluctuation determinants, and $x_{cl}\left( \tau \right) $
is the instanton solution. The prime on the det indicates that the zero
eigenvalue is to omitted, and $\omega $ is the frequency of harmonic
amplitude oscillations in the wells about the minima. The general formulas
were presented to evaluate the preexponential factor in the tunnel splitting
or the tunneling rate.\cite{23,24} It was found that what is need for this
evaluation is the asymptotic $\left( \tau \rightarrow \pm \infty \right) $
behavior of the instanton velocity, 
\begin{equation}
\frac{dx_{cl}}{d\tau }\approx a\exp \left( \mp \omega \tau \right) ,\text{
as }\tau \rightarrow \pm \infty .  \eqnum{11}
\end{equation}
Then the ground-state tunnel splitting is\cite{23,24} 
\begin{equation}
\Delta E_0=2\left| a\right| \left( \frac{m\omega }\pi \right) ^{1/2}\exp
\left( -W_0\right) .  \eqnum{12}
\end{equation}
In most physical applications, however, the ground-state tunnel splitting
can be best estimated as 
\begin{equation}
\Delta E_0=p_0\omega \left( \frac{W_0}{2\pi }\right) ^{1/2}\exp \left(
-W_0\right) ,  \eqnum{13}
\end{equation}
where the dimensionless prefactor $p_0$ is often relevant to experiments. It
is noted that Eq. (12) is based on quantum tunneling at the level of ground
state, and the temperature dependence of the tunneling frequency (i.e.,
tunneling at excited levels) is not taken into account. The instanton
technique is suitable only for the evaluation of the tunneling rate or the
tunnel splitting at the vacuum level, since the usual (vacuum) instantons
satisfy the vacuum boundary conditions. Recently, different types of
pseudoparticle configurations (periodic or nonvacuum instantons) are found
which satisfy periodic boundary conditions.\cite{25}

For the same tunneling problem, the WKB answer for the splittings of the $n$%
th degenerate excited levels or the imaginary parts of the metastable levels
is\cite{26,24} 
\begin{equation}
\Delta E_n\left( \text{or }%
\mathop{\rm Im}%
E_n\right) =\frac{\omega \left( E_n\right) }\pi \exp \left( -W\right) , 
\eqnum{14}
\end{equation}
where $\omega \left( E_n\right) =\frac{2\pi }{t\left( E_n\right) }$ is the
energy-dependent frequency. $t\left( E_n\right) $ is the period of the
real-time oscillation in the potential well, 
\begin{equation}
t\left( E_n\right) =\sqrt{2m}\int_{x_1\left( E_n\right) }^{x_2\left(
E_n\right) }\frac{dx}{\sqrt{E_n-U\left( x\right) }},  \eqnum{15}
\end{equation}
where $x_{1,2}\left( E_n\right) $ are the turning points for the particle
oscillating inside the potential $U\left( x\right) $. The
functional-integral and the WKB\ method showed that for the potentials
parabolic near the bottom the result Eq. (14) should be multiplied by $\sqrt{%
\frac \pi e}\frac{\left( 2n+1\right) ^{n+1/2}}{2^ne^nn!}$.\cite{27} This
factor is very close to 1 for all $n$: 1.075 for $n=0$, 1.028 for $n=1$,
1.017 for $n=2$, etc. Stirling's formula for $n!$ shows that this factor
trends to 1 as $n\rightarrow \infty $. Therefore, this correction factor,
however, does not change much in front of the exponentially small action
term in Eq. (14). It is noted that Eq. (14) can be obtained by using the
connection formulas near a linear turning point,\cite{26} or by matching the
WKB wavefunction in the classically forbidden region to the exact harmonic
oscillator wavefunction near the classical turning point.\cite{24} It was
shown that the WKB method is equivalent to the instanton method for the
ground-state tunnel splitting.\cite{24} Liang {\it et al.} showed that the
WKB\ method and the periodic instanton method give the same result for the
excited-state tunnel splitting.\cite{25}

For the present problem, we find that $\Delta \varepsilon _A=\frac 2{%
t_A\left( E\right) }\exp \left( -W_A\right) $, where $t_A\left( E\right) =%
\frac 2{\omega _1\left( E\right) }K\left( k_1^{\prime }\right) $ and $%
k_1^{\prime }=\sqrt{1-k_1^2}$. The same method can be applied to the
instanton $B$ passing through the large barrier. And the result is $\Delta
\varepsilon _B=\frac 2{t_B\left( E\right) }\exp \left( -W_B\right) $ for $%
0<E<U_S$, where $t_B\left( E\right) =t_A\left( E\right) $, and $W_B$ has the
same expression as Eq. (8) but taking $\xi _1$ as $\xi _2=\frac{1+\beta }{%
1-\beta }$. For $U_S\leq E\leq U_L$, the imaginary part of the metastable
level is $%
\mathop{\rm Im}%
E=\frac 2{\widetilde{t}_B\left( E\right) }\exp \left( -2\widetilde{W}%
_B\right) $, where $\widetilde{t}_B\left( E\right) =\frac 2{\omega _2\left(
E\right) }K\left( k_2^{\prime }\right) $, 
\begin{equation}
\widetilde{W}_B=2m\xi _3\left( 1+\alpha \right) \omega _2\left[ \frac 1{%
k_2^2-\xi _3}E\left( k_2\right) -\frac 1{\xi _3}K\left( k_2\right) +\frac{%
k_2^2+\xi _3^2+2k_2^2\xi _3}{\xi _3}\Pi \left( k_2,\xi _3\right) \right] , 
\eqnum{16}
\end{equation}
with $k_2^{\prime }=\sqrt{1-k_2^2}$, $\omega _2=\frac{\omega _0}{g_2}$, $g_2=%
\sqrt{\frac 2{\alpha -\beta }}$, $\xi _3=\frac{1+\beta }{\alpha -\beta }$,
and $k_2^2=\frac{\left( \alpha -1\right) \left( 1+\beta \right) }{2\left(
\alpha -\beta \right) }$. Now we discuss the low energy limit of the level
splitting. With the help of harmonic oscillator approximation for energy
near the bottom of the potential well, $\varepsilon _n=\left( n+\frac 12%
\right) \Omega $, and 
\[
\Omega =\sqrt{\frac 1m\left( \frac{d^2U}{d\phi ^2}\right) _{\phi =\phi _0}}=%
\sqrt{\frac{2K_{\Vert }V}{\hbar m}}\sin \theta _0\cos \phi _0,
\]
$W_{A\left( B\right) }$ can be expanded as 
\begin{eqnarray}
W_{A\left( B\right) ,n} &=&W_{A\left( B\right) ,0}-\left( n+\frac 12\right)
+\left( n+\frac 12\right) \ln \left( \frac{n+1/2}{8\sqrt{\lambda }S\sin
\theta _0\cos ^3\phi _0}\right) ,  \eqnum{17a} \\
W_{A\left( B\right) ,0} &=&2\sqrt{\lambda }S\sin \theta _0\left( \cos \phi
_0\mp 2\sin \phi _0\arcsin \sqrt{\frac{1\mp \sin \phi _0}2}\right) , 
\eqnum{17b}
\end{eqnarray}
where ``$-$'' for the instanton $A$, and ``$+$'' for the instanton $B$.
Therefore, the tunnel splittings for instantons $A$ and $B$ are 
\begin{equation}
\hbar \Delta \varepsilon _{A\left( B\right) ,n}=\frac{2^{3/2}}{\sqrt{\pi }n!}%
\left( K_{\bot }V\right) \sqrt{\lambda }S^{-1}\sin \theta _0\cos \phi
_0\left( 8\sqrt{\lambda }S\sin \theta _0\cos ^3\phi _0\right) ^{n+1/2}\exp
\left( -W_{A\left( B\right) ,0}\right) .  \eqnum{18}
\end{equation}
Equation (17b) shows that the WKB exponent for instanton $A$ is smaller than
that for instanton $B$ at finite magnetic field because the barrier through
which instanton $B$ must tunnel is higher than that for instanton $A$. For
the case of ground-state resonance, i.e., $n=0$, Eq. (18) reduces to 
\begin{equation}
\Delta \varepsilon _{A\left( B\right) ,0}=\frac 4{\sqrt{\pi }}\lambda
^{1/4}S^{1/2}\left( \sin \theta _0\cos ^3\phi _0\right) ^{1/2}\Omega \exp
\left( -W_{A\left( B\right) ,0}\right) .  \eqnum{19a}
\end{equation}
It is not particularly illuminating to write out the prefactor $p_0$ (see
Eq. (13)), although it can be seen that it is dimensionless and independent
of the volume $V$ of the particle. Note that Eq. (13) is the approximate
formula for ground-state tunnel splitting with exponential accuracy. In most
physical applications, the preexponential factor can be best estimated as an
attempt frequency. The apparent disagreement with Eq. (13) was also found in
other spin-tunneling problems.\cite{24} However, in the case of zero
magnetic field Eq. (19a) reduces to 
\begin{eqnarray}
\Delta \varepsilon _{A,0}\left( H=0\right)  &=&\Delta \varepsilon
_{B,0}\left( H=0\right)   \nonumber \\
&=&4\Omega _0\left( \frac{W_0}{2\pi }\right) ^{1/2}\exp \left( -W_0\right) ,
\eqnum{19b}
\end{eqnarray}
where $\Omega _0=\Omega \left( H=0\right) $, and $W_0=W_{A,0}\left(
H=0\right) =W_{B,0}\left( H=0\right) =2\sqrt{\lambda }S$. Compared with Eq.
(13), $p_0=4$ for this case.

It is noted that $\hbar \Delta \varepsilon _{A\left( B\right) ,n}$ is only
the level shift induced by tunneling between degenerate excited states
through a single barrier. $U\left( \phi \right) $ can be regarded as a
one-dimensional superlattice with the sublattices $A$ and $B$. The Bloch
states for sublattices $A$ and $B$ are 
\begin{equation}
\Phi _A\left( \xi ,\phi \right) =\frac 1{\sqrt{L}}\sum_ne^{i\xi \phi
_n}\varphi _A\left( \phi -\phi _n\right) ,  \eqnum{20a}
\end{equation}
and 
\begin{equation}
\Phi _B\left( \xi ,\phi \right) =\frac 1{\sqrt{L}}\sum_ne^{i\xi \left( \phi
_n+a\right) }\varphi _B\left( \phi -\phi _n-a\right) ,  \eqnum{20b}
\end{equation}
where $\phi _n=2n\pi +\phi _0$, $L=N\left( a+b\right) $, $a=\pi -2\phi _0$,
and $b=\pi +2\phi _0$. The total wavefunction is 
\begin{equation}
\Psi _\xi \left( \phi \right) =a_A\left( \xi \right) \Phi _A\left( \xi ,\phi
\right) +a_B\left( \xi \right) \Phi _B\left( \xi ,\phi \right) .  \eqnum{21}
\end{equation}
Including the Wess-Zumino phase, we derive the secular equation in the
tight-binding approximation as 
\begin{equation}
\left[ 
\begin{array}{ll}
\varepsilon _n-E\left( \xi \right) & e^{i\left( \xi -\mu \right) a}\Delta
\varepsilon _{A,n}+e^{-i\left( \xi -\mu \right) b}\Delta \varepsilon _{B,n}
\\ 
e^{-i\left( \xi -\mu \right) a}\Delta \varepsilon _{A,n}+e^{i\left( \xi -\mu
\right) b}\Delta \varepsilon _{B,n} & \varepsilon _n-E\left( \xi \right)
\end{array}
\right] \left[ 
\begin{array}{l}
a_A\left( \xi \right) \\ 
a_B\left( \xi \right)
\end{array}
\right] =0,  \eqnum{22}
\end{equation}
where $\mu =S\left( 1-\cos \theta _0\right) $, and the Bloch wave vector $%
\xi =0$ in the first Brillouin zone. Then the tunnel splitting of the $n$th
excited level is 
\begin{equation}
\Delta \varepsilon _n=2\sqrt{\left( \Delta \varepsilon _{A,n}\right)
^2+\left( \Delta \varepsilon _{B,n}\right) ^2+2\left( \Delta \varepsilon
_{A,n}\right) \left( \Delta \varepsilon _{B,n}\right) \cos \left[ 2\pi
S\left( 1-\cos \theta _0\right) \right] }.  \eqnum{23}
\end{equation}
Another approach to obtain Eq. (23) is to calculate the transition amplitude
directly. The subtle point in evaluating the transition amplitude is how to
arrange the instantons and anti-instantons appropriately to satisfy the
boundary condition. Note that one configuration starting from $\left| \theta
_0,\phi _0\right\rangle $ and ending at $\left| \theta _0,\phi
_0\right\rangle $ can be an arbitrary permutation of the pairs $\left(
AB\right) $, $\left( AA^{-}\right) $, $\left( B^{-}B\right) $, and $\left(
B^{-}A^{-}\right) $. Introducing $s$, $t$, $p$, and $q$ as the numbers of $A$%
, $B$, $A^{-}$, and $B^{-}$ in the instantons and anti-instantons pairs, and 
$i$, $j$, $k$, and $l$ as those of $\left( AB\right) $, $\left(
AA^{-}\right) $, $\left( B^{-}B\right) $, and $\left( B^{-}A^{-}\right) $,
we have 
\begin{equation}
i+j=s,i+k=t,j+l=p,k+l=q.  \eqnum{24}
\end{equation}
Therefore, $s+q=t+p$, and only three variables are independent. Now the
transition amplitude is 
\begin{eqnarray}
K_E &=&\sqrt{\frac \Omega {\pi \hbar }}e^{-\left( n+\frac 12\right) \Omega
}\sum_{s,t,p,q}\frac{N\left( s,t,p,q\right) }{\left( s+t+p+q\right) !}\left[
\left( \hbar \Delta \varepsilon _{A,n}T\right) ^{s+p}\left( \hbar \Delta
\varepsilon _{B,n}T\right) ^{t+q}\right.  \nonumber \\
&&\left. \times e^{-iS\left( 1-\cos \theta _0\right) \left( \pi -2\phi
_0\right) \left( s-p\right) }e^{-iS\left( 1-\cos \theta _0\right) \left( \pi
+2\phi _0\right) \left( t-q\right) }\right] \delta _{s+q,t+p},  \eqnum{25}
\end{eqnarray}
where $N\left( s,t,p,q\right) $ represents the number of different
configurations for a given set of $\left\{ s,t,p,q\right\} $. It can be
calculated as 
\begin{equation}
N\left( s,t,p,q\right) =\sum_{i=\max \left\{ 0,t-q\right\} }^{\min \left\{
s,t\right\} }\frac{\left( i+j+k+l\right) !}{i!j!k!l!},  \eqnum{26}
\end{equation}
which has a simple result 
\begin{equation}
N\left( s,t,p,q\right) =\frac{\left[ \left( s+q\right) !\right] ^2}{s!t!p!q!}%
.  \eqnum{27}
\end{equation}
Then we have 
\begin{equation}
K_E=\sqrt{\frac \Omega {\pi \hbar }}e^{-\left( n+\frac 12\right) \Omega
}\cosh \left[ \sqrt{\left( \Delta \varepsilon _{A,n}\right) ^2+\left( \Delta
\varepsilon _{B,n}\right) ^2+2\left( \Delta \varepsilon _{A,n}\right) \left(
\Delta \varepsilon _{B,n}\right) \cos \left[ 2\pi S\left( 1-\cos \theta
_0\right) \right] }T\right] ,  \eqnum{28}
\end{equation}
and we can read off the splitting of the $n$th excited level shown in Eq.
(23) directly.

Finally we discuss the phase transition from classical to quantum behavior.
It was found that for a particle moving in a double-well potential $U\left(
x\right) $, the behavior of the energy-dependent period of oscillations $%
P\left( E\right) $ in the Euclidean potential $-U\left( x\right) $
determines the order of the quantum-classical transition.\cite{28} If $%
P\left( E\right) $ monotonically increases with the amplitude of
oscillations, i.e., with decreasing energy $E$, the transition is of second
order. The corresponding crossover temperature is $T_0^{\left( 2\right) }=%
\frac{\widetilde{\omega }_0}{2\pi }$, $\widetilde{\omega }_0=\sqrt{\frac 1m%
\left| U^{\prime \prime }\left( x_{sad}\right) \right| }$ is the frequency
of small oscillations near the bottom of $-U\left( x\right) $, where $%
x_{sad} $ corresponds to the top (the saddle point) of the barrier. If,
however, the dependence of $P\left( E\right) $ is non-monotonic, the first
order crossover takes place. For this case, the period of the instanton $A$
is $P_A\left( E\right) =\frac 4{\omega _1}K\left( k_1\right) $. The
monotonically decreasing behavior of $P_A\left( E\right) $ in the domain $%
0\leq E\leq U_S$ is shown in Fig. 1 for several values of $\sin \phi _0$,
which shows that the second-order phase transition takes place, and the
crossover temperature is 
\begin{equation}
k_BT_{0,A}^{\left( 2\right) }=\frac{\sqrt{K_{\Vert }K_{\bot }}V}{\pi \hbar S}%
\sin \theta _0\sqrt{1-\sin \phi _0}.  \eqnum{29a}
\end{equation}
The period of the instanton $B$ is $P_B\left( E\right) =P_A\left( E\right) $
for $0\leq E\leq U_S$, while $P_B\left( E\right) =\frac 4{\omega _2}K\left(
k_2\right) $ for $U_S\leq E\leq U_L$. Fig. 2 shows the monotonically
decreasing behavior of $P_B\left( E\right) $ for the whole domain $0\leq
E\leq U_L$. Again one finds a second-order transition from the thermal to
quantum regime, and the crossover temperature is 
\begin{equation}
k_BT_{0,B}^{\left( 2\right) }=\frac{\sqrt{K_{\Vert }K_{\bot }}V}{\pi \hbar S}%
\sin \theta _0\sqrt{1+\sin \phi _0}.  \eqnum{29b}
\end{equation}

At zero magnetic field, $\Delta \varepsilon _{A,n}=\Delta \varepsilon _{B,n}$%
, the tunnel splitting is suppressed to zero for the half-integer total
spins by the destructive interfering Wess-Zumino-Berry phases, which is in
good agreement with the Kramers theorem. The presence of a magnetic field
perpendicular to the plane of rotation of magnetization yields an additional
contribution to the Berry phase, resulting constructive and destructive
interferences alternatively for both integer and half-integer spins.
Tunneling is thus periodically suppressed. At finite magnetic field, the
low-lying tunneling level spectrum depends on the parity of total spins. If $%
S$ is an integer, the tunnel splitting is 
\begin{equation}
\Delta \varepsilon _n=2\sqrt{\left( \Delta \varepsilon _{A,n}\right)
^2+\left( \Delta \varepsilon _{B,n}\right) ^2+2\left( \Delta \varepsilon
_{A,n}\right) \left( \Delta \varepsilon _{B,n}\right) \cos \left( 2\pi S\cos
\theta _0\right) }.  \eqnum{30a}
\end{equation}
While if $S$ is a half-integer, the splitting is 
\begin{equation}
\Delta \varepsilon _n=2\sqrt{\left( \Delta \varepsilon _{A,n}\right)
^2+\left( \Delta \varepsilon _{B,n}\right) ^2-2\left( \Delta \varepsilon
_{A,n}\right) \left( \Delta \varepsilon _{B,n}\right) \cos \left( 2\pi S\cos
\theta _0\right) }.  \eqnum{30b}
\end{equation}
Equations (30a) and (30b) clearly show oscillations of the tunnel splitting
as a function of the transverse magnetic field, together with the parity
effect. Note that the magnetic field along the medium axis does not produce
any oscillations. As $\Delta \varepsilon _{A,n}>\Delta \varepsilon _{B,n}$
at finite magnetic field, the tunnel splitting will not be suppressed to
zero even if the total spin is a half-integer. The similar oscillation of
tunnel splitting with the magnetic field was also found in Ref. 29, while
the system considered in Ref. 29 is a single-domain antiferromagnetic
particle with a field along the hard axis. The most interesting observation
is that only the $\widehat{z}$ component of the magnetic field (i.e., along
the hard axis) can lead to the oscillation of the tunnel splitting. As shown
in Fig. 3, the splitting depends on the magnitude of the field ${\bf H}$ and
its angle $\theta _H$ with the hard axis, and even a small misalignment of
the field with the $\widehat{z}$ axis can completely destroy the oscillation
effect. For small $\theta _H$ the tunnel splitting oscillates with the
field, whereas no oscillation is shown up for large $\theta _H$. In the
latter case, a much stronger increase of tunnel splitting with the field is
shown. This strong dependence on the orientation of the field can be
observed for ground-state resonance as well as excited-state resonance. As a
result, we conclude that {\it the topological phase interference or
spin-parity effects depend on the orientation of the external magnetic field
distinctly}. {\it The distinct angular dependence, together with the
oscillation of the tunnel splittings with the field, may provide an
independent experimental test for the spin phase interference effects in FM
particles}. This ``Aharonov-Bohm'' type of oscillation in magnetic system is
analogous to the oscillations as a function of external flux in a SQUID
ring. Due to the topological nature of the Berry phase, these spin-parity
effects are independent of details such as the magnitude of total spins and
the shape of the soliton. The transition from classical to quantum behavior
is also studied for the small and large tunneling barrier, respectively. By
calculating the periods of instanton $A$ and $B$ analytically, we find the
monotonically decreasing behavior of the periods with increasing energy,
which yields a second-order phase transition. Our results should be useful
for a quantitative understanding on the spin-parity effects and the
quantum-classical crossover for FM particles in an arbitrarily directed
magnetic field. It is noted that the theoretical results presented here are
based on the instanton method, which is semiclassical in nature, i.e., valid
for large spins and in the continuum limit. The theoretical calculations
performed in this paper can be extended to the single-domain
antiferromagnetic particles where the relevant quantity is the excess spin
due to the small noncompensation of two sublattices. Work along this line is
still in progress. We hope that the theoretical results obtained in the
present work will stimulate more experiments whose aim is observing the
topological phase interference effects and the quantum-classical transition
in resonant quantum tunneling of magnetization in nanometer-scale
single-domain ferromagnets.


\begin{references}
\bibitem{1}  For a review, see {\it Quantum Tunneling of Magnetization},
edited by L. Gunther and B. Barbara (Kluwer, Dordrecht, 1995); and E. M.
Chudnovsky and J. Tejada, {\it Macroscopic Quantum Tunneling of the Magnetic
Moment} (Cambridge University Press, 1997).

\bibitem{2}  E. Fradkin, {\it Field Theories of Condensed Matter Systems}
(Addison-Wesley, Reading, MA, 1992), Chap. 5.

\bibitem{3}  D. Loss, D. P. DiVincenzo, and G. Grinstein, Phys. Rev. Lett. 
{\bf 69}, 3232 (1992); J. von Delft and G. L. Henley, Phys. Rev. Lett. {\bf %
69}, 3236 (1992).

\bibitem{4}  A. Garg, Europhys. Lett. {\bf 22}, 205 (1993); cond-mat/9906203.

\bibitem{5}  H. B. Braun and D. Loss, Europhys. Lett. {\bf 31}, 555 (1995).

\bibitem{6}  S. E. Barnes, Advances in Physics {\bf 30}, 801 (1981); J.
Phys.: Condens. Matter {\bf 6}, 719 (1994); {\bf 10}, L665 (1998);
cond-mat/9710302; cond-mat/9901219; cond-mat/9907257; S. E. Barnes, R.
Ballou, and J. Strelen, Phys. Rev. Lett. {\bf 79}, 289 (1997).

\bibitem{7}  E. N. Bogachek and I. V. Krive, Phys. Rev. B {\bf 46}, 14559
(1992).

\bibitem{8}  E. M. Chudnovsky and D. P. DiVincenzo, Phys. Rev. B {\bf 48},
10548 (1993); A. Garg, Phys. Rev. B {\bf 51}, 15161 (1995).

\bibitem{9}  X. -B. Wang and F. -C. Pu, J. Phys.: Condens. Matter {\bf 8},
L541 (1996); {\bf 9}, 693 (1997).

\bibitem{10}  E. M. Chudnovsky, J. Magn. Magn. Mater. {\bf 140-144}, 1821
(1995); V. Yu. Golyshev and A. F. Popkov, Europhys. Lett. {\bf 29}, 327
(1995); A. Chiolero and D. Loss, Phys. Rev. Lett. {\bf 80}, 169 (1998).

\bibitem{11}  J. M. Duan and A. Garg, Physica B {\bf 194-196}, 323 (1994);
J. Phys.: Condens. Matter {\bf 7}, 2171 (1995).

\bibitem{12}  H. B. Braun and D. Loss, Phys. Rev. B {\bf 53}, 3237 (1996).

\bibitem{13}  B. A. Ivanov, A. K. Kolezhuk, and V. E. Kireev, Phys. Rev. B 
{\bf 58}, 11514 (1998).

\bibitem{14}  Rong L\"{u} {\it et al.}, Euro. Phys. J B{\bf 3},35 (1998);
Phys. Rev. B {\bf 58}, 8542 (1998); Jia-Lin Zhu {\it et al.}, Eur. Phys. J. 
{\bf B4}, 223 (1998); Rong L\"{u} {\it et al.}, Phys. Rev. B {\bf 60}, 3534
(1999).

\bibitem{15}  W. Wernsdorfer and R. Sessoli, Science {\bf 284}\-, 133 (1999).

\bibitem{16}  N. V. Prokof'ev and P. C. E. Stamp, J. Low Temp. Phys. {\bf 104%
}, 143 (1996); {\it ib id }{\bf 113}, 1147 (1998); G. Rose and P. C. E.
Stamp, {\it ib id }{\bf 113}, 1147 (1998); L. Gunther, Europhys. Lett. {\bf %
39}, 1 (1997).

\bibitem{17}  D. A. Garanin and E. M. Chudnovsky, Phys. Rev. B {\bf 56},
11102 (1997).

\bibitem{18}  M. N. Leuenberger and D. A. Garanin, cond-mat/9810156.

\bibitem{19}  A. Garg, Phys. Rev. Lett. {\bf 83}, 4385 (1999);
cond-mat/9907197; cond-mat/9907198; Phys. Rev. Lett. {\bf 81}, 1513 (1998).

\bibitem{20}  I. Chiorescu {\it et al.}, cond-mat/9910117; W. Wernsdorfer 
{\it et al.}, cond-mat/0001346.

\bibitem{21}  J. -Q. Liang {\it et al.}, cond-mat/0001142.

\bibitem{22}  N. V. Prokof'ev and P. C. E. Stamp, Phys. Rev. Lett. {\bf 80},
5794 (1998); J. Phys.: Condens. Matter {\bf 5}, L663 (1998).

\bibitem{23}  S. Coleman, Phys. Rev. B {\bf 15}, 2929 (1977); {\it Aspects
of Symmetry} (Cambridge University Press, Cambridge, England), Chap. 7.

\bibitem{24}  A. Garg and G. -H. Kim, Phys. Rev. B {\bf 45}, 12921 (1992);
A. Garg, cond-mat/0003115.

\bibitem{25}  J. -Q. Liang {\it et al.}, Phys. Rev. B {\bf 57}, 529 (1998);
J. -Q. Liang and H. J. W. M\"{u}ller-Kirsten, Phys. Rev. D {\bf 46}, 4685
(1992).

\bibitem{26}  L. D. Landau and E. M. Lifshitz, {\it Quantum Mechanics}
(Pergamon, New York, 1977), Chap. VII, Sec. 50, Problem 3.

\bibitem{27}  U. Weiss and W. Haeffner, Phys. Rev. D {\bf 27}, 2916 (1983);
H. K. Shepard, Phys. Rev. D {\bf 27}, 1288 (1983).

\bibitem{28}  E. M. Chudnovsky and D. A. Garanin, Phys. Rev. Lett. {\bf 79},
4469 (1997); D. A. Garanin, X. M. Hidalgo, and E. M. Chudnovsky, Phys. Rev.
B {\bf 57}, 13639 (1998).

\bibitem{29}  Y. -H. Nie {\it et al}., J. Phys.: Condens. Matter {\bf 12},
L87 (2000).

Figure Captions:

Fig. 1 The relative period $P_A\left( E\right) /P_A\left( E=U_S\right) $ of
the periodic instanton $A$ as a function of energy $E/K_{\Vert }V\sin
^2\theta _0$ in the domain $0\leq E\leq U_S$.

Fig. 2 The relative period $P_B\left( E\right) /P_B\left( E=U_S\right) $ of
the periodic instanton $B$ as a function of energy $E/K_{\Vert }V\sin
^2\theta _0$ in the domain $0\leq E\leq U_L$.

Fig. 3 The relative tunnel splitting of the ground level $\left( n=0\right) $
$\Delta \varepsilon _0/\Delta \varepsilon _{B,0}\left( H=0\right) $ as a
function of $H/H_0$ for $\theta _H=0{{}^{\circ }}$, $1{{}^{\circ }}$, $3{%
{}^{\circ }}$, $5{{}^{\circ }}$, and $90{{}^{\circ }}$. Here $\lambda
=K_{\Vert }/K_{\bot }=0.1$.
\end{references}
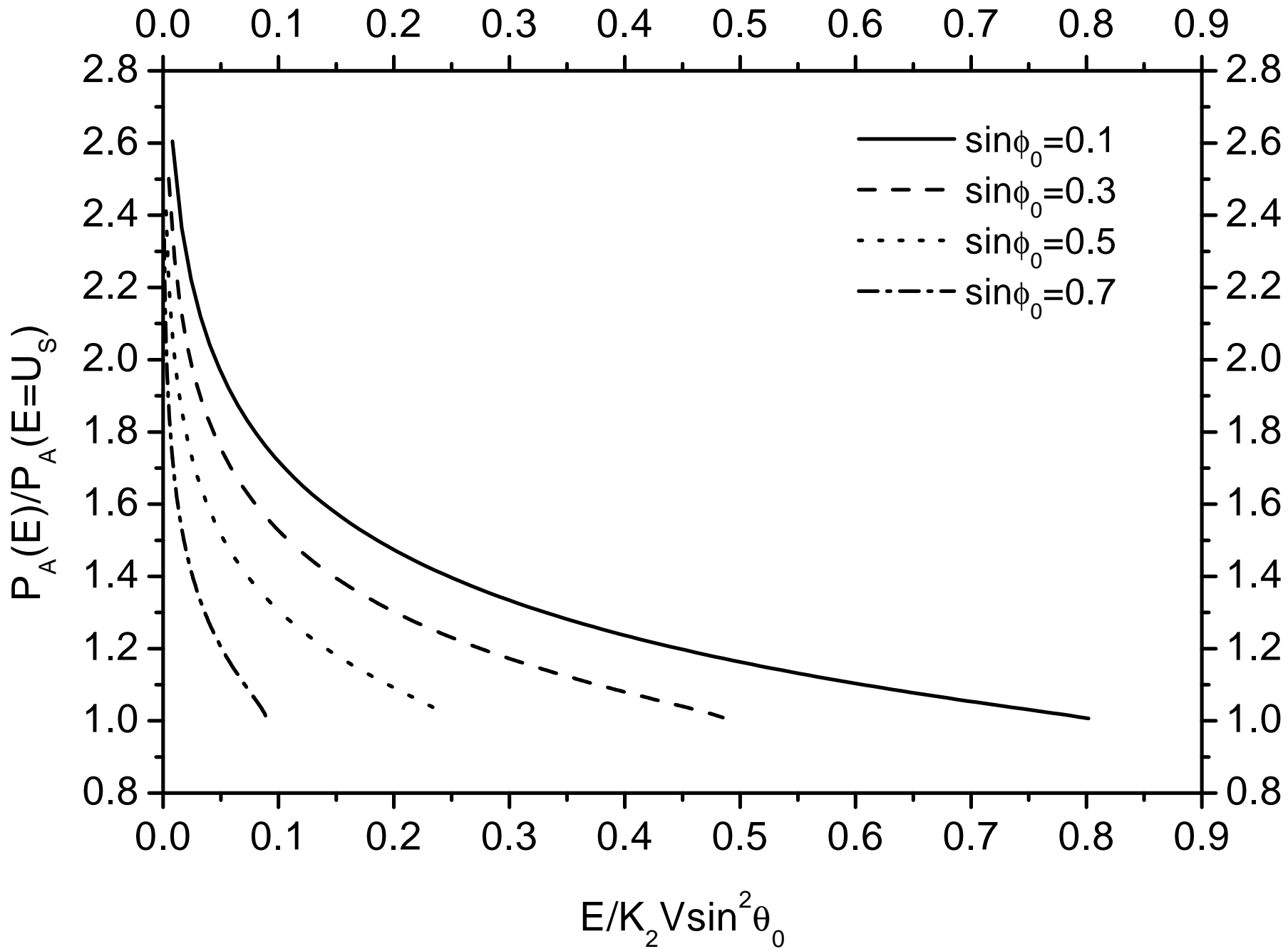
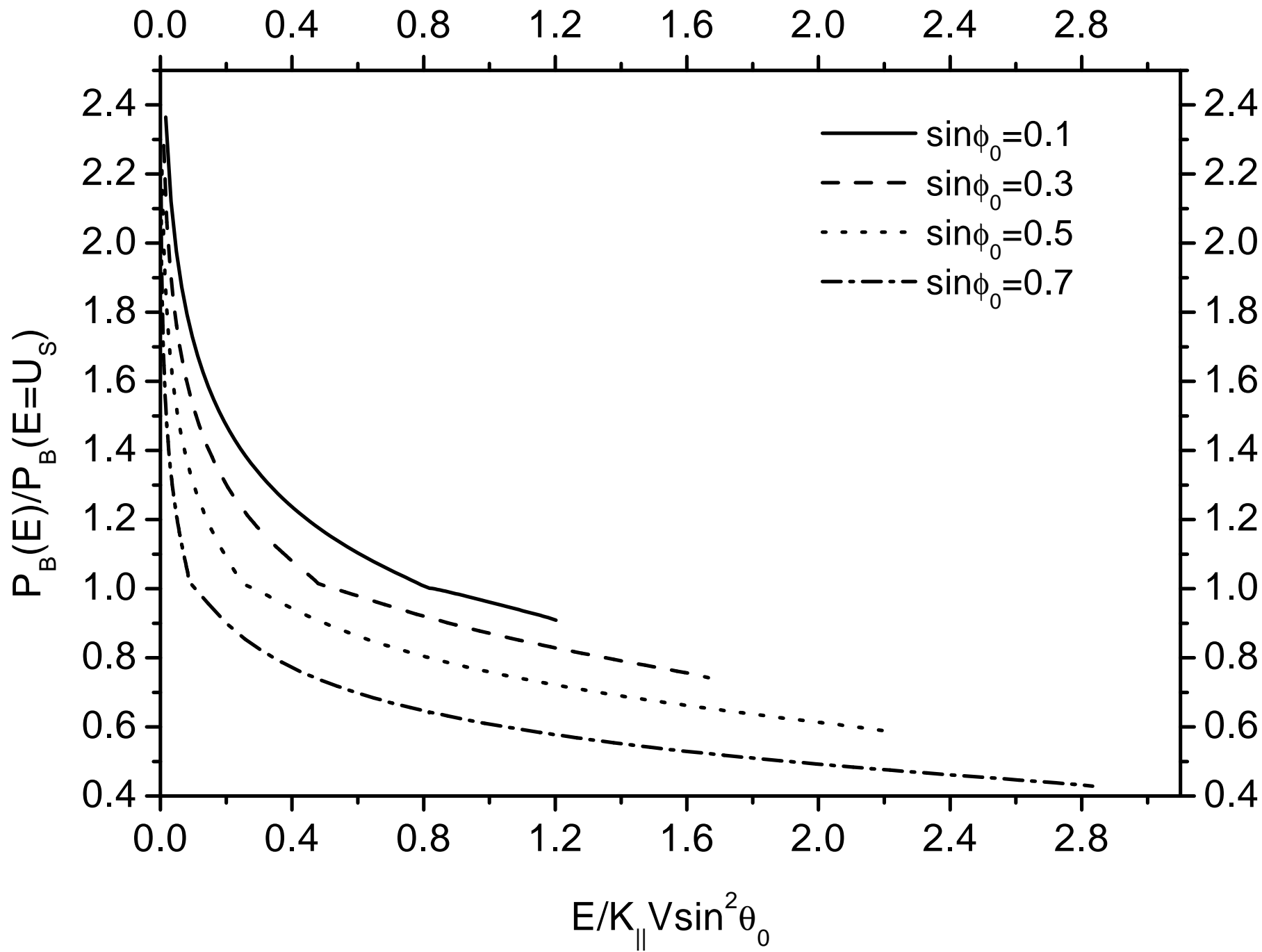
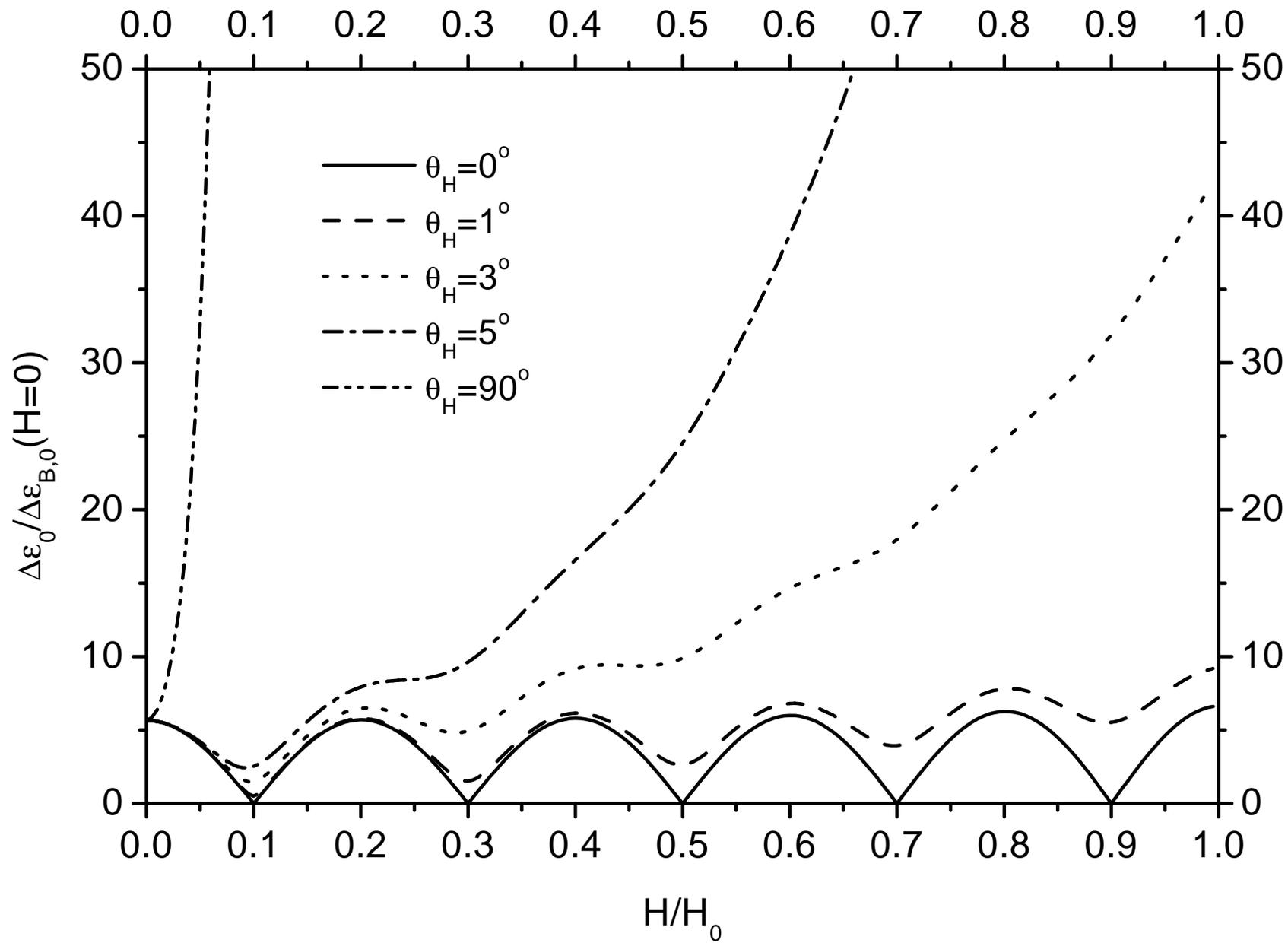
\end{document}